\documentclass[twocolumn,showpacs,preprintnumbers,amsmath,amssymb]{revtex4}

\usepackage{graphicx}
\usepackage{dcolumn}
\usepackage{bm}
\usepackage{epsfig}

\begin{document}

\title[Electron Trapping]{Electron trapping by electric field reversal and Fermi mechanism}
\author{M\'{a}rio J. Pinheiro}
\address{Department of Physics and Centro de Fisica de Plasmas, Instituto Superior T\'{e}cnico,
Av. Rovisco Pais, 1049-001 Lisboa, Portugal}

\email{mpinheiro@ist.utl.pt}

\homepage{http://alfa.ist.utl.pt/~pinheiro}

\pacs{52.10.+y, 52.20.Fs, 52.25.Dg, 52.25.Gj, 52.27.Aj}
\date{\today}

\begin{abstract}
We investigate the existence of the electric field reversal in the
negative glow of a dc discharge, its location, the width of the
well trapping the electrons, the slow electrons scattering time,
and as well the trapping time. Based on a stress-energy tensor
analysis we show the inherent instability of the well. We suggest
that the Fermi mechanism is a possible process for pumping out
electrons from the through, and linking this phenomena with
electrostatic plasma instabilities. A power law distribution
function for trapped electrons is also obtained. Analytical
expressions are derived which can be used to calculate these
characteristics from geometrical dimensions and the operational
parameters of the discharge.

\end{abstract}

\pacs{52.10.+y, 52.20.Fs, 52.25.Dg, 52.25.Gj, 52.27.Aj}

\bibliographystyle{apsrev}

\maketitle


\section{Introduction}

The phenomena of field reversal of the axial electric field in the
negative glow of a dc discharge is of great importance, since the
fraction of ions returning to the cathode depends on its existence
and location. Technological application of gas discharges,
particularly to plasma display panels, needs a better knowledge of
the processes involved. The study of nonlocal phenomena in
electron kinetics of collisional gas discharge plasma have shown
that in the presence of field reversals the bulk electrons in the
cathode plasma are clearly separated in two groups of slow
electrons: trapped and free electrons\cite{Kolobov}. Trapped
electrons give no contribution to the current but represent the
majority of the electron population.

The first field reversal it was shown qualitatively to be located
near the end of the negative glow (NG) where the plasma density
attains the greatest magnitude. If the discharge length is enough,
it appears a second field reversal on the boundary between the
Faraday dark space and the positive column. Also, it was shown in
the previously referred theoretical work that ions produced to the
left of this first reversal location move to the cathode by
ambipolar diffusion and ions generated to the right of this
location drift to the anode. For a review see also~\cite{Godyak}.
Those characteristic were experimentally observed by laser
optogalvanic spectroscopy~\cite{Gottscho}.

Boeuf {\it et al}~\cite{Boeuf} with a simple fluid model gave an
analytical expression of the field reversal location which showed
to depend solely on the cathode sheath length, the gap length, and
the ionization relaxation length. They obtained as well a simple
analytical expression giving the fraction of ions returning to the
cathode and the magnitude of the plasma maximum density.

In the present Letter we introduce a quite simple dielectric-like
model of a plasma-sheath system. This approach have been addressed
by other authors~\cite{Taillet69, Harmon76} to explain how the
electrical field inversion occurs at the interface between the
plasma sheath and the beginning of the negative glow. The aim of
this Letter is to obtain more information about the fundamental
properties related to field inversion phenomena in the frame of a
dielectric model. It is obtained a simple analytical dependence of
the axial location where field reversal occurs in terms of
macroscopic parameters. In addition, it is obtained the magnitude
of the minimum electric field inside the through, the trapped well
length, and the trapping time of the slow electrons into the well.
We emphasize in particular the description of the dielectric
behavior and do not contemplate plasma chemistry and
plasma-surface interactions.

The analytical results hereby obtained could be useful for hybrid
fluid-particle models (e.g., Fiala {\it et al.}~\cite{Boeuf94}),
since simple criteria can be applied to accurately remove
electrons from the simulations.

On the ground of the stress-energy tensor considerations it is
shown the inherent instability of the field inversion sheath. The
slow electrons distribution function is obtained assuming the
Fermi~\cite{Fermi} mechanism responsible for their acceleration
from the trapping well.

\section{Theoretical model}

Lets consider a plasma formed between two parallel-plate
electrodes due to an applied dc electric field. We assume a planar
geometry, but extension to cylindrical geometry is
straightforward. The applied voltage is $V_a$ and we assume the
cathode fall length is $l$ and the negative glow + eventually the
positive column extends over the length $l_0$, such that the total
length is $L=l + l_0$. We have
\begin{equation}\label{Eq1}
-V_a = l E_s + l_0 E_p,
\end{equation}
where $E_s$ and $E_p$ are, resp., the electric fields in the
sheath and NG (possibly including the positive column).

At the end of the cathode sheath it must be verified the following
boundary condition by the displacement field $\mathbf{D}$
\begin{equation}\label{Eq2}
\mathbf{n}.(\mathbf{D}_p - \mathbf{D}_s) = \sigma.
\end{equation}
Here, $\sigma$ is the surface charge density accumulated at the
boundary surface and $\mathbf{n}$ is the normal to the surface. In
more explicit form,
\begin{equation}\label{Eq3}
\varepsilon_p E_p - \varepsilon_s E_s = \sigma.
\end{equation}
Here, $\varepsilon_s$ and $\varepsilon_p$ are, resp., the
electrical permittivity of the sheath and the positive column. We
have to solve the following algebraic system of equations
\begin{equation}\label{Eq5}
\begin{array}{cc}
  l_0 E_p + l E_s & = - V_a, \\
  \varepsilon_p E_p - \varepsilon_s E_s & = \sigma. \\
\end{array}
\end{equation}
They give the electric field strength in each region
\begin{equation}\label{Eq6}
\begin{array}{cc}
  E_s =  & -\frac{V_a}{L} \left(1-\alpha + \frac{l_o \sigma}{V_a \varepsilon_s}\right)\frac{1}{1-\frac{l\alpha}{L}}, \\
  E_p =  & -\frac{V_a}{L} \left(1-\frac{l \sigma}{V_a \varepsilon_s}  \right) \frac{1}{1-\frac{l\alpha}{L}}.\\
\end{array}
\end{equation}
Here, we define
$\alpha=1-\frac{\varepsilon_p}{\varepsilon_s}=\frac{\omega_p^2}{\nu_{en}^2}$.
Recall that in DC case,
$\varepsilon_p=1-\frac{\omega_p^2}{\nu_{en}^2}$, and
$\varepsilon_s=\varepsilon_0$, with $\omega_p$ denoting the plasma
frequency and $\nu_{en}$ the electron-neutral collision frequency.
In fact, our assumption $\varepsilon_s=\varepsilon_0$ is plainly
justified, since experiments have shown the occurrence of a
significant gas heating and a corresponding gas density reduction
in the cathode fall region, mainly due to symmetric charge
exchanges processes which lead to an efficient conversion of
electrical energy to heavy-particle kinetic energy and thus to
heating~\cite{Hartog88}.

 Two extreme cases can be considered: {\bf i}) $\omega_p
> \nu_{en}$, implying $\varepsilon_p < 0$, meaning that
$\tau_{coll} > \tau_{plasma}$, i.e, non-collisional regime
prevails; {\bf ii}) $\omega_p < \nu_{en}$, $\varepsilon_p > 0$,
and then $\tau_{coll}
> \tau_{plasma}$, i.e, collisional regime dominates.

From the above Eqs.~\ref{Eq6} we estimate the field inversion
should occurs for the condition $1-\frac{l_o \alpha}{L}=0$, which
give the position on the axis where field inversion occurs:
\begin{equation}\label{Eq8}
\frac{l_o}{L} = \frac{\nu_{en}^2}{\omega_p^2}.
\end{equation}

From Eq.~\ref{Eq8} we can resume a criteria for field reversal: it
only occurs in the non-collisional regime; by the contrary, in the
collisional regime and to the extent of validity of this simple
model, no field reversal will occur, since the slow electrons
scattering time inside the well is higher than the the well
lifetime, and collisions (in particular, coulombian collisions)
and trapping become competitive processes. A similar condition was
obtained in~\cite{Nishikawa69} when studying the effect of
electron trapping in ion-wave instability. Likewise, a
self-consistent analytic model~\cite{Kolobov} have shown that at
at sufficiently high pressure, field reversal is absent.

Due to the accumulation of slow electrons after a distance
$\xi_c=L-l_0$, real charges accumulated on a surface separating
the cathode fall region from the negative glow. Naturally, it
appears polarization charges on each side of this surface and a
double layer is created with a surface charge $-\sigma_1' < 0$ on
the cathode side and $\sigma_2'$ on the anode side. But, $\sigma'
= (\mathbf{P} \cdot \mathbf{n})$,
$\mathbf{P}=\mathbf{\mathbf{\varepsilon}_0} \chi_e \mathbf{E}$
with $\varepsilon = \varepsilon_0 (1 + \chi_e)$, $\chi_e$ denoting
the dimensionless quantity called electric susceptibility. As the
electric displacement is the same everywhere, we have
$\mathbf{D}_0 = \mathbf{D}_1 = \mathbf{D}_2$. Thus, the residual
(true) surface charge in between is given by
\begin{equation}\label{Eq9}
\sigma = - \sigma_1' + \sigma_2'.
\end{equation}
After a straightforward but lengthy algebraic operation we obtain
\begin{equation}\label{Eq10}
\sigma = \varepsilon_p V_a \frac{B}{A},
\end{equation}
where
\begin{equation}\label{Eq11}
A = L \left( -1+ \frac{\varepsilon_0 -
\varepsilon_s}{\varepsilon_p} \right) + l\left( -
\frac{\varepsilon_p}{\varepsilon_s} +
\frac{\varepsilon_s}{\varepsilon_p} \right),
\end{equation}
and
\begin{equation}\label{Eq12}
B = \frac{\varepsilon_0 (\varepsilon_s - \varepsilon_p)}{
\varepsilon_s \varepsilon_p }.
\end{equation}
We can verify that $\sigma$ must be equal to
\begin{equation}\label{Eqsig}
\sigma = \alpha \frac{V_a \varepsilon_0}{2 l_0}.
\end{equation}
Considering that $\sigma = \varepsilon_0 \chi_e E$, we determine
the minimum value of the electric field at the reversal point:
\begin{equation}\label{Eq13}
E_m = \frac{\omega_p^2}{\nu_{en}^2} \frac{V_a}{2 l_0 \chi_e}.
\end{equation}

Here, $\chi_e=\varepsilon_{rw}-1$, with $\varepsilon_{rw}$
designating the relative permittivity of the plasma trapped in the
well. From the above equation we can obtain a more practical
expression for the electrical field at its minimum strength
\begin{equation}\label{Eq14}
E_m = -
\frac{n_{ep}}{n_{ew}}\frac{\nu_{enw}^2}{\nu_{en}^2}\frac{V_a}{e
l_0} \approx - \frac{n_{ep}}{n_{ew}} \frac{T_{ew}}{T_{ep}}
\frac{V_a}{2 l_0}.
\end{equation}
The magnitude of the minimum electric field depends on the length
of the negative glow $l_0$. This also means that without NG there
is no place for field reversal, and also the bigger the length the
minor the electric field. The length of the negative glow can be
estimated by the free path length $l_0$ of the fastest electrons
possessing an energy equal to the cathode potential fall value
$eV_a$:
\begin{equation}\label{}
l_0 = \int_{0}^{eV_a} \frac{d w}{(N F(w))}.
\end{equation}
Here, $w$ is the electrons kinetic energy and $NF(w)$ is the
stopping power. For example, for He, it is estimated $p l_0=0.02
eV_a$ ~\cite{Kolobov} (in cm.Torr units, with $V_a$ in Volt). We
denote by $n_{ew}$ the density of trapped electrons and by
$T_{ew}$ their respective temperature. Altogether, $n_{ep}$ and
$T_{ep}$ are, resp., the electron density and electron temperature
in the negative glow region.

By other side, we can estimate the true surface charge density
accumulated on the interface of the two regions by the expression
\begin{equation}\label{Eq15}
\sigma = \frac{Q}{A} = - \frac{n_{ep} e A \Delta \xi}{A}.
\end{equation}
Here, $Q$ is the total charge over the cross sectional area where
the current flows and $\Delta \xi$ is the width of the potential
well.

\subsection{Instability and width of the potential well}

From Eqs.~\ref{Eqsig} and ~\ref{Eq15} it is easily obtained the
trapping well width
\begin{equation}\label{Eq16}
\Delta \xi = - \frac{e V_a}{2 m l_0 \nu_{enw}^2}.
\end{equation}
It is expected that the potential trough should have a
characteristic width of the order in between the electron Debye
length ($\lambda_{De}=\sqrt{\frac{\varepsilon_0 kT_e}{n_e e^2}}$)
and the mean scattering length. Using Eq.~\ref{Eq16}, in a He
plasma and assuming $V_a=1$ kV, $l_0=1$ m and $\nu_{en}=1.85
\times 10^{9}$ s$^{-1}$ (with $T_e=0.03$ eV) at 1 Torr ($n=3.22
\times 10^{16}$ cm$^{-3}$) we estimate $\Delta \xi \approx 2.6
\times 10^{-3}$ cm, while the Debye length is $\lambda_{De}=2.4
\times 10^{-3}$ cm. So, our Eq.~\ref{Eq16} gives a good order of
magnitude for the potential width, which is expected to be in fact
of the same order of magnitude than the Debye length.

Table I present the set of parameters used to obtain our
estimations. We give in Table II the estimate of the minimum
electric field attained inside the well. The first field reversal
at $\xi_c \approx l_{NG}$ corresponds to the maximum density
$n_{ew} \gg n_{ep}$~\cite{Boeuf,Tsendin2001}. So, the assumed
values for the ratio of electron temperatures and densities of the
trapped electrons and electrons on the NG are typical estimates.

\begin{table}
  \centering
  \caption{Data used for $E/p=100$ V$/$cm$/$Torr. Cross sections and electron
  temperatures are
  taken from Siglo Data base, CPAT and Kinema Software, http://www.Siglo-Kinema.com }\label{Table1}
  \begin{tabular}{|c|c|c|}
    \hline
    Gas & $T_e$ (eV) & $\sigma$ ($10^{-16}$ cm$^{2}$)\\
    \hline
    Ar & 8 & 4.0 \\
    He & 35 & 2.0 \\
    O$_2$ & 6 & 4.5 \\
    N$_2$ & 4 & 9.0 \\
    H$_2$ & 8 & 6.0 \\
    \hline
  \end{tabular}
\end{table}

It can be shown that there is no finite configuration of fields
and plasma that can be in equilibrium without some external
stress~\cite{Longmire}. Consequently, this trough is dim to be
unstable and burst electrons periodically (or in a chaotic
process), releasing the trapped electrons to the main plasma. This
phenomena produces local perturbation in the ionization rate and
the electric field giving rise to ionization waves (striations).
In the next section, we will calculate the time of trapping with a
simple Brownian model.

\begin{table}\label{Table2}
  \centering
  \caption{Minimum electric field at reversal point and well width.
  Conditions: He gas, $p=1$ Torr, $l_0=20$ cm, $V_a=1$ kV, $\frac{T_{ew}}{T_{ep}}=0.1$, $\frac{n_{ew}}{n_{ep}}=10$.}\label{Table1}
  \begin{tabular}{|c|c|}
    \hline 
    $E_m$ (V.cm$^{-1}$) & $\Delta \xi$ (cm) \\
    \hline
    $\lesssim -2.5$  & $2.6 \times 10^{-3}$ \\
    \hline
  \end{tabular}
\end{table}

From Eq.~\ref{Eq8} we calculate the cathode fall length for some
gases. For this purpose we took He and H$_2$ data as reference for
atomic and molecular gases, resp. The orders of magnitude are the
same, with the exception of Ar. Due to Ramsauer effect direct
comparison is difficult.

In Table III it is shown a comparison of the experimental cathode
fall distances to the theoretical prediction, as given by
Eq.~\ref{Eq16}. Taking into account the limitations of this model
these estimates are well consistent with experimental
data~\cite{Brown59}.

\begin{table}
  \centering
  \caption{Comparison between theoretical and experimental cathode fall distance at p=1 Torr,
  $E/p$=100 V$/$cm$/$Torr. Experimental data are collected from Ref.~\cite{Brown59}. }\label{Table2}
\begin{tabular}{|c|c|c|}
  \hline
  Gas & $\xi_c^{teo}$ (cm) & $\xi_c^{exp}$ (cm) \\
  \hline
  Ar    & 7.40      & 0.29 (Al)  \\
  He    & 1.32     & 1.32 (Al) \\
  $H_2$ & 0.80      & 0.80  (Cu) \\
  $N_2$ & 0.45     & 0.31 (Al) \\
  $Ne$  & 0.80      & 0.64 (Al) \\
  $O_2$ & 0.30      & 0.24 (Al) \\
  \hline
\end{tabular}
\end{table}

\subsection{Lifetime of a slow electron in the potential well}

The trapped electrons most probably diffuse inside the well with a
characteristic time much shorter than the lifetime of the through.
Trapping can be avoided by Coulomb collisions~\cite{Nishikawa69}
or by the ion-wave instability, both probably one outcome of the
stress energy unbalance as previously mentioned. We consider a
simple Brownian motion model for the slow electrons to obtain the
scattering time $\tau$, and the lifetime T of the well. A
Fermi-like model will allow us to obtain the slow electron energy
distribution function.

Considering the slow electron jiggling within the well, the
estimated scattering time is
\begin{equation}\label{scat1}
\tau = \frac{(\Delta \xi)^2}{ \mathcal{D}_e }.
\end{equation}
Here, $\mathcal{D}_e$ is the electron diffusion coefficient at
thermal velocities.

\begin{table*}
  \caption{Scattering time and trapping time in the well. The parameters are:
  $E/N=100$ Td, $T_g=300$ K, $V_a=1$ kV and $l_0=0.1$ m.}\label{Table 4}
  \begin{ruledtabular}
  \begin{tabular}{cccccc}
    Gas & $\mathcal{D}_e$ (cm$^2$.s$^{-1}$)\footnote{Data obtained through resolution of the homogeneous electron Boltzmann equation
    with two term expansion of the distribution function in spherical harmonics, M. J. Pinheiro and J. Loureiro,
    J. Phys. D.: Appl. Phys. {\bf 35} 1 (2002) }  & $\nu_{enw} (s^{-1})$ \footnote{Same remark as in ${}^a$} & $\Delta \xi (cm)$ & $\tau$ (s) & T (s) \\
    \hline
    Ar    & $2.52 \times 10^6$  & $8.10 \times 10^9$ & $1.34 \times 10^{-3}$  & $7.10 \times 10^{-13}$ & $3.97 \times 10^{-5}$ \\
    He    & $5.99 \times 10^6$  & $2.39 \times 10^9$ & $1.54 \times 10^{-2}$ & $3.95 \times 10^{-11}$ & $1.70 \times 10^{-5}$\\
    N$_2$ & $6.11 \times 10^5$  & $6.15 \times 10^9$ & $2.32 \times 10^{-3}$ & $8.81 \times 10^{-12}$  &$1.64 \times 10^{-4}$  \\
    CO$_2$ & $1.70 \times 10^6$ & $3.60 \times 10^9$ & $6.78 \times
    10^{-3}$& $2.70 \times 10^{-11}$ & $5.90 \times 10^{-5}$ \\
  \end{tabular}
  \end{ruledtabular}
\end{table*}

The fluctuations arising in the plasma are due to the breaking of
the well and we can estimate the amplitude of the fluctuating
field by means of Eq.~\ref{Eq14}. We obtain
\begin{equation}\label{}
\delta E_m =
\frac{n_{ep}}{n_{ew}}\frac{\nu_{enw}^2}{\nu_{en}^2}\frac{V_a}{e
l_0^2} \Delta \xi.
\end{equation}

Then, we have
\begin{equation}\label{}
\mathcal{E}_c = \frac{\delta E_m}{E_m} = \frac{\Delta \xi}{l_0}.
\end{equation}

In Table IV we summarize scattering and trapping times for a few
gases.

\subsection{Power-law slow electrons distribution function}

As slow electrons are trapped by the electric field inversion,
some process must be at work to pull them out from the well. We
suggest that fluctuations of the electric field in the plasma
(with order of magnitude of $\mathcal{E}_c$)act over electrons
giving energy to the slow ones, which collide with those
irregularities as with heavy particles. From this mechanism it
results a gain of energy as well a loss. This model was first
advanced by E. Fermi~\cite{Fermi} when developing a theory of the
origin of cosmic radiation. We shall focus here on the rate at
which energy is acquired.

The average energy gain per collision by the trapped electrons (in
order of magnitude) is given by
\begin{equation}\label{}
\Delta w = \overline{U} w(t),
\end{equation}
with $\overline{U}\cong \mathcal{E}_c^2$ and where $w$ is their
kinetic energy. After $N$ collisions the electrons energy will be
\begin{equation}\label{}
w(t) = \varepsilon_{t} \exp \left( \frac{\overline{U}t}{\tau}
\right),
\end{equation}
with $\varepsilon_t$ being their thermal energy, typical of slow
electrons.
\begin{figure}
  \includegraphics[width=3 in, height=4.5 in]{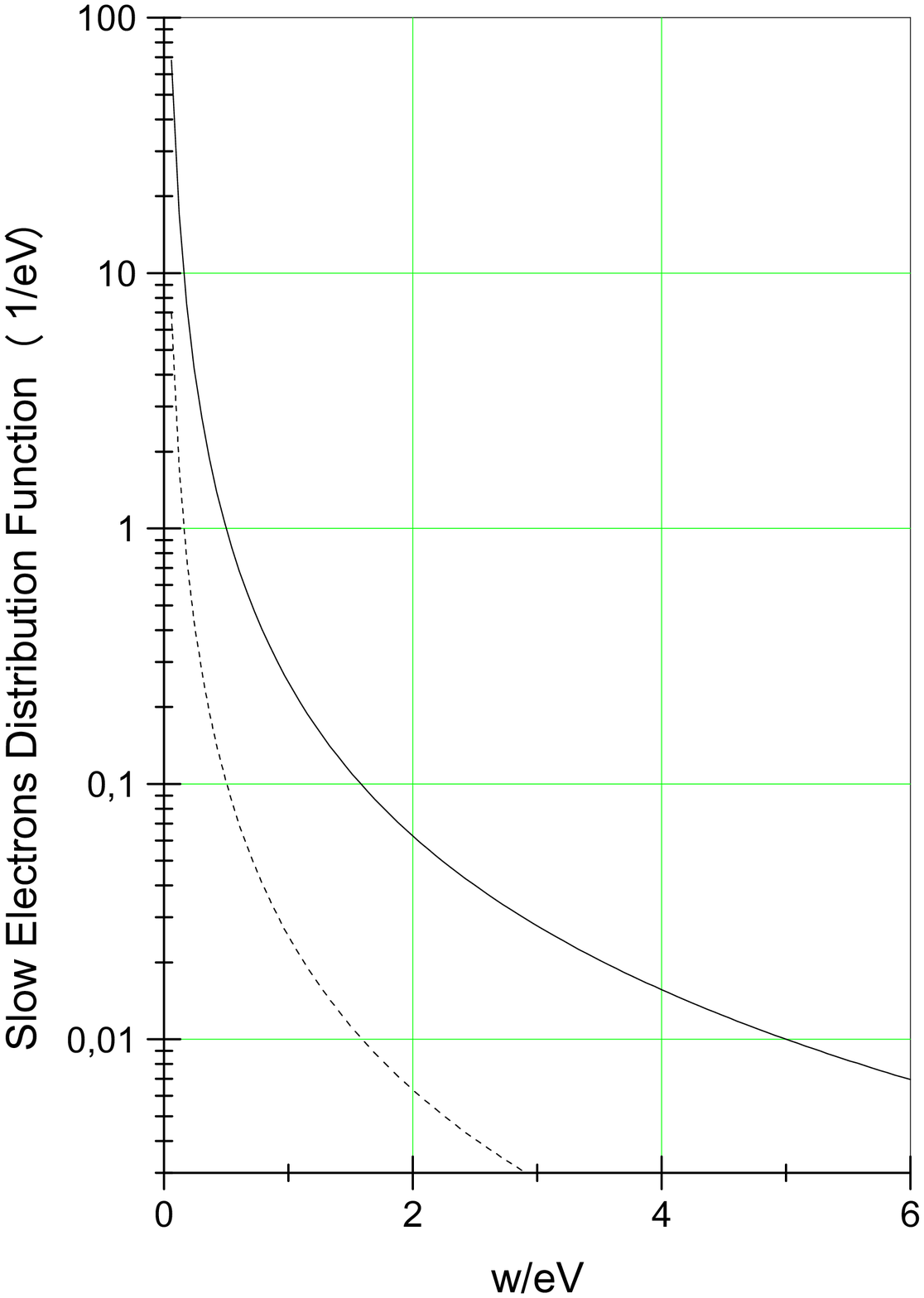}\\
  \caption{Slow electrons distribution function vs. energy, for the same conditions as presented in Table IV. Solid curve: Ar, broken curve: N$_2$.}\label{}
\end{figure}
The time between scattering collisions is $\tau$. Assuming a
Poisson distribution $P(t)$ for electrons escaping from the
trapping, then we state
\begin{equation}\label{5}
  P(t)=\exp(-t/\tau)dt/T.
\end{equation}
The probability distribution of the energy gained is a function of
one random variable (the energy), such as
\begin{equation}\label{6}
  f_w(w)d w = P\{ w<\bar{w}<w+dw \}.
\end{equation}
This density $f_w(w)$ can be determined in terms of the density
P(t). Denoting by $t_1=T$ the real root of the equation
$w=w(t_1=T)$, then it can be readily shown that slow electrons
obey in fact to the following power-law distribution function
\begin{equation}\label{7}
  f_w(w) d w = \frac{\tau}{\bar{U}T}
  \varepsilon_{t}^{\frac{\tau}{\bar{U}T}} \frac{d
  w}{w^{1+\tau/\bar{U}T}}.
\end{equation}
Like many man made and naturally occurring phenomena (e.g.,
earthquakes magnitude, distribution of income), it is expected the
trapped electron distribution function to be a power-law (see
Eq.~\ref{7}), hence $1 + \frac{\tau}{\mathcal{E}_c^2 T} = n$, with
$n=2 \div 4$ as a reasonable guess. Hence, we estimate the
trapping time to be
\begin{equation}\label{traptime}
T \approx \frac{\tau}{\mathcal{E}_c^2 n}.
\end{equation}

Fig.1 shows the slow electrons distribution function pumped out
from the well for two cases: Ar (solid curve), and N$_2$ (broken
curve). It was chosen a power exponent $n=2$. Those distributions
show that the higher confining time is associated with less slow
electrons present in the well. When the width of the well
increases (from solid to broken curve) the scattering time become
longer, and as well the confining time, due to a decrease of the
relative number of slow electrons per given energy. This mechanism
of pumping out of slow (trapped) electrons from the well can
possibly explains the generation of electrostatic plasma
instabilities.

Note that the trapping time is, in fact, proportional to the
length of the NG and inversely proportional to the electrons
diffusion coefficient at thermal energies:
\begin{equation}\label{}
T \approx \frac{l_0^2}{\mathcal{D}_e}.
\end{equation}
The survival frequency of trapped electrons is $\nu_t=1/T$. As the
electrons diffusion coefficient are typically higher in atomic
gases, it is natural to expect plasma instabilities and waves with
higher frequencies in atomic gases. This result is in agreement
with a kinetic analysis of instabilities in microwave
discharges~\cite{Tatarova1}. In addition, the length of the NG
will influence the magnitude of the frequencies registered by the
instabilities, since wavelengths have more or less space to
build-up. Table~\ref{Table 4} summarizes the previous results for
some atomic and molecular gases. The transport parameters used
therefor where calculated by solving the electron Boltzmann
equation, under the two-term approximation, in a steady-state
Townsend discharge~\cite{Pinheiro}

\section{Conclusion}

We have shown in the framework of a simple dielectric model that
the magnitude of the minimum electric field (on the edge of the
negative glow) depends directly on the applied voltage and is
inversely proportional to the NG length.

The width of the well trapping the slow electrons is directly
dependent on the applied electric field and is inversely
proportional to the square of the electron-neutral collision
frequency for slow electrons. It is, as well, inversely
proportional to the NG length, and has typically the extension of
a Debye length. We state that for typical conditions of a
low-pressure glow-discharge, field reversal occurs whenever
$\omega_p
> \nu_{en}$, due to a lack of collisions necessary to pump out electrons from the well.
Furthermore, the analytical expressions obtained for the
scattering and trapping time of the slow electrons are potentially
useful in hybrid fluid-particle plasma modelling.


\end{document}